\documentclass[reprint,
superscriptaddress,
 amsmath,
 amssymb,
 aps,
 prd,
floatfix,
]{revtex4-2}

\usepackage{graphicx}% Include figure files
\usepackage{dcolumn}% Align table columns on decimal point
\usepackage{bm}% bold math
\usepackage{hyperref}% add hypertext capabilities
\usepackage{aas_macros}
\usepackage{textcomp}
\usepackage{xfrac}
\usepackage{comment}
\usepackage{multirow}   % multirow entries in tables

\usepackage{makecell}
\usepackage{xspace}
\usepackage[font=small,labelfont=bf,
   justification=justified,
   format=plain]{caption}
\usepackage{xcolor}

\newcommand{\lya}{Lyman-$\alpha$\xspace}
\newcommand{\dF}{\delta_F}
\newcommand{\kl}{\kappa_{{\rm Ly}\alpha}}
\newcommand{\kc}{\kappa_{\rm CMB}}

\newcommand{\vth}{\bm{\theta}}
\newcommand{\Mpch}{\,h^{-1}{\rm Mpc}}
\newcommand{\rperp}{r_\perp}
\newcommand{\rpar}{r_\parallel}

\providecommand{\Nside}{N_{\rm side}}

\begin{document}

\preprint{arXiv:xxxx.xxxx}

\title{Weak Lensing Measurements from the \lya Forest}

\author{An\v{z}e Slosar}
 \affiliation{Physics Department, Brookhaven National Laboratory, Upton, NY 11973}

\date{\today}% It is always \today, today,

\begin{abstract}
We attempt to measure the weak lensing of the \lya forest by projecting a quadratic estimator of the forest-forest and quasar-forest pair correlations, built from the DESI DR1 data, onto deflection templates constructed from low-redshift galaxies and quasars from DESI and BOSS. After validating deflection templates against the ACT DR6 and Planck PR4 CMB lensing maps and testing the estimator with injected displacements we measure the lensing amplitude $A_L$ on the data. Consistent with predictions in the literature, the current generation of public data does not have sufficient signal-to-noise to measure this effect unequivocally: we find the lensing amplitude $A_L=0.42\pm0.43$ from the forest auto-correlation and $0.87\pm0.48$ from the quasar--forest cross-correlation. These two nearly independent measurements combine to $A_L=0.62\pm0.33$, consistent with both the fiducial value $A_L=1$ and with no lensing. 
\end{abstract}

\maketitle

\section{Introduction}

The effect of weak gravitational lensing is to transform the sky in a brightness-preserving manner: with respect to the unlensed sky, the images are shifted and stretched \citep{astro-ph/9912508}. For galaxy images this leads to magnification and shear of the objects \citep{1411.0115}; for the cosmic microwave background it leads to local variations in the power spectrum \citep{astro-ph/0601594}. The same effect also applies to the \lya forest \citep{1706.07870,1706.08939,2005.04109,2209.04564,2410.20014}: the quasar sight-lines are displaced and their relative distances are modified with respect to the unlensed case. However, pixel pair correlations remain those of the physical separation, not the apparent separation: this allows one to estimate the lensing potential. The estimator essentially measures the local correlation function, compares it with the sky-mean one and thus deduces how the sightlines have been pushed around. It is quadratic in the \lya flux fluctuations, analogous to the CMB case where the convergence map is quadratic in the CMB temperature fluctuations.

The literature to date has mostly focused on lensing as traced by the forest-forest fluctuations and forecasted that the signal might be detectable in the cross-correlation with the low redshift sample using complete DESI \lya forest at around 4 sigma \cite{2410.20014}.  Despite the final DESI release not being available, the work to date indicates that it might be possible to get non-trivial measurements even with the currently available Data Release 1 (DR1) of the DESI data. This is the main motivation for this paper.

An important innovation is that we consider the quasar-forest cross-correlations in addition to forest-forest cross-correlations. In BAO studies with \lya forest these add the same level of signal than auto-correlations so we can expect similar improvements here.

As noted before, the most promising way in the low SNR regime is to attempt this measurement with a high fidelity tracer of the lensing field and we chose the lensing potential traced by the galaxies at the low redshift. In effect this measurement performs a detection in the three-point functions $\delta_t \delta_F \delta_F$ and $\delta_t \delta_F \delta_Q$ where $\delta_t$ is a tracer at some low redshift. In practice, we use multiple samples to generate low-redshift templates of displacement which we cross-correlate with pairs in the forest region.

It might also be possible to measure this effect by cross-correlating with the CMB lensing map $\kc$. However, the bispectrum $\kc \delta_F \delta_F$ contains contributions from both gravitational lensing and the density bispectrum: $\kc$ traces density fluctuations at the redshift of the \lya forest as well, and the forest has a non-zero bispectrum with them \citep{astro-ph/0302112,astro-ph/0308151,1701.03375}. This effect is analogous to the intrinsic alignments in galaxy lensing \citep{1407.6990} and therefore has to be modeled or subtracted. We leave this for future work.

Finally, we note that lensing also smears the \lya forest correlation function in the transverse direction, analogous to the smoothing of the CMB acoustic peaks. This is another effect that is likely to have a negligible impact on the current data but should be considered for the future.

This paper is structured as follows: in Section \ref{sec:theory} we lay down the theory and the estimator basics. In Section \ref{sec:templates} we build the deflection templates from DESI and BOSS tracers and validate them against CMB lensing. In Section \ref{sec:data} we describe the \lya forest data and the response kernels, in Section \ref{sec:results} we validate the estimator with injected signal and present the measurement and its robustness, and we conclude in Section \ref{sec:conclusions}.

Throughout the paper we work with a fixed flat $\Lambda$CDM cosmological model close to the Planck 2018 best fit \citep{1807.06209}: $H_0=67.66\,{\rm km\,s^{-1}\,Mpc^{-1}}$, $\Omega_bh^2=0.02242$, $\Omega_ch^2=0.11933$, $n_s=0.9665$, $A_s=2.105\times10^{-9}$ and one massive neutrino with $m_\nu=0.06$\,eV, with distances, the linear growth factor and the linear and non-linear (HMcode-2020) matter power spectra computed with CAMB \citep{Lewis:1999bs}. All separations are comoving and quoted in $\Mpch$.

\section{Theory}
\label{sec:theory}

Let $\delta_F$ denote the usual \lya forest flux fluctuation
\begin{equation}
    \delta_F(\hat{n},\chi) = \frac{F(\hat{n},\chi)- \bar{F}}{\bar{F}},
\end{equation}
where $\hat{n}$ denotes the position on the sky and $\chi$ the comoving distance (inferred from the sightline position and the pixel redshift).

For the quasar sample the overdensity $\delta_Q$ is defined analogously, but note that quasars are point tracers rather than a field sampled along sightlines.

\subsection{Lensing of $\delta_F$ pair correlations}
Let $\dF(\vth,\chi)$ be the forest flux fluctuation observed along a sightline at angular position $\vth$ and comoving distance $\chi$. Lensing maps the observed position to the true one, $\vth_{\rm true}=\vth+\bm\alpha(\vth)$, with $\bm\alpha=\nabla\phi$ and convergence $\kappa=-\nabla^2\phi/2$, where $\phi$ is the lensing potential.  The observed field is the true field sampled at $\vth+\bm\alpha$. For a pair of sightlines $a,b$ the correlation of two pixels at transverse separation $r_\perp=\chi|\vth_a-\vth_b|$ and radial separation $r_\parallel$ is the true correlation at the displaced separation,
\begin{multline}
    \langle\dF(a,\chi_p)\dF(b,\chi_q)\rangle=\xi_F\!\left(\bigl|\vth_{ab}+\bm\alpha_a-\bm\alpha_b\bigr|\chi,\,r_\parallel\right) \\
\simeq\xi_F(r_\perp,r_\parallel)+\frac{\partial\xi_F}{\partial r_\perp}\,\chi\,\hat\vth_{ab}\cdot(\bm\alpha_a-\bm\alpha_b) \\
 = \xi_F(r_\perp,r_\parallel)+G_{ab}(\alpha) \xi_F',
\label{eq:pair}
\end{multline}
with $\hat\vth_{ab}$ the unit pair direction (from $b$ to $a$), and where we have defined the shorthand $G_{ab}(\alpha)\equiv\chi\,\hat\vth_{ab}\cdot(\bm\alpha_a-\bm\alpha_b)$ for the geometric factor and $\xi_F' = {\partial\xi_F}/{\partial r_\perp}$ for the response kernel. For a pair separation small compared with the wavelength of the lens, $\bm\alpha_a-\bm\alpha_b\simeq(\vth_{ab}\cdot\nabla)\nabla\phi$: the convergence dilates the transverse correlation isotropically and the shear makes it anisotropic.

\subsection{Lensing of $\delta_Q-\delta_F$ cross-correlations}

The quasar--forest cross-correlation is lensed in an analogous way, with one important difference: quasar overdensities are also subject to magnification bias. Under lensing, the observed quasar density transforms as
\begin{equation}
    \delta_Q(\vth) \rightarrow \delta_Q(\vth+\bm\alpha) + (5s-2)\kappa(\vth), 
\end{equation}
where $s={\rm d}\log_{10}N(<m)/{\rm d}m$ is the slope of the cumulative number counts of the quasar sample at its flux limit. However, because the $\kappa$ field does not correlate with the $\delta_F$ field (they are at two different redshifts), magnification bias does not enter the $\delta_Q$--$\delta_F$ estimator at first order. In other words, the two point function between low-redshift lensing templates and $\delta_Q$ would tease out these correlations, but the three point function also involving $\delta_F$ only teases them out at quadratic order.

Therefore the cross-pair equivalent of Equation \ref{eq:pair} is, for a quasar $q$ at $(\vth_q,\chi_q)$ and a forest pixel $p$ at $(\vth_p,\chi_p)$ on another sightline,
\begin{multline}
\langle\dF(p)\,|\,\text{quasar at }q\rangle=\xi_{qF}\!\left(\bigl|\vth_{qp}+\bm\alpha_q-\bm\alpha_p\bigr|\chi,\,r_\parallel\right)\\
\simeq\xi_{qF}(r_\perp,r_\parallel)+G_{qp}(\alpha)\,\xi'_{qF},\qquad r_\parallel=\chi_p-\chi_q,
\label{eq:pairqf}
\end{multline}
with the same geometric factor $G_{qp}(\alpha)=\chi\,\hat\vth_{qp}\cdot(\bm\alpha_q-\bm\alpha_p)$ and $\xi'_{qF}=\partial\xi_{qF}/\partial r_\perp$. The only notable difference is that the quasar--forest correlation is computed by stacking $\dF$ around quasar positions, so its radial separation is signed and the quasar itself counts with unit weight.

\subsection{Quadratic estimator with a deflection template}
\label{sec:qe}
Let us assume that the total deflection can be written as a sum over $N_t$ deflection template maps with amplitudes $A_i$ 
\begin{equation}
   \bm\alpha(\vth) = \sum_i A_i \bm\alpha_i^T(\vth)
\end{equation}
We now use a matched-filter approach to estimate the amount of the $\bm\alpha_i^T$ component in the data.  Given the $i$-th component, consider the quantity
\begin{multline}
\label{eq:qe}
\Delta q_i=\sum_{ab}\;G_{ab}(\alpha_i) \times\\
\sum_{p\in a,q\in b}w_pw_q\,\left(\dF(p)\dF(q)-\xi_F(p,q)\right)\,\xi'_F(r_\perp,r_\parallel).
\end{multline}
Its expectation value is 
\begin{multline}
    \left<\Delta q_i\right> =  \sum_{abj}\; 
\sum_{p\in a,q\in b}w_pw_q\, G_{ab}(\alpha_i) G_{ab}(\alpha_j) \xi'^2_F(r_\perp,r_\parallel) A_j\\
= F_{ij} A_j
\end{multline}
and therefore the estimate of the amplitude factors is given by
\begin{equation}
\hat A=F^{-1}\Delta q,\qquad    
\end{equation}
Here $F$ is the $N_t \times N_t$ response matrix and $A$ is the $N_t$ vector of template amplitudes. The subtraction of $\xi_F(p,q)$ inside $\Delta q_i$ subtracts the mean field: it removes the correlation of the sightline density (and hence of the pair geometry) with the template, which does not vanish on a masked footprint. %Errors are estimated by a jackknife over HEALPix regions of the pair midpoints (nside 8, about 300 regions) and checked against random templates (Section~\ref{sec:results}).

Here $w_p$ and $w_q$ are diagonal pixel weights. Since the response weighting is already contained in $\xi_F'$, these are simply the diagonal inverse-variance weights $w_p = 1/(\sigma_n^2 + \sigma_i^2)$, where $\sigma_n^2$ is the noise variance of the pixel and $\sigma_i^2$ the intrinsic variance of the flux field. %a full per-sightline inverse covariance would shape the weights along the line of sight and we estimate its gain at 8 per cent in signal-to-noise, not worth its complexity here.

The formula for quasar-forest cross-correlations is exactly the same, except that the sum instead of being over all pixels pairs is over pairs of nearby quasars and forest pixels.

The estimator requires the templates to be evaluated at the sight-line positions, but the choice of templates is up to us. For example, the templates could be individual modes of the convergence plane, which would allow one to reconstruct the map. In our case we generate them from low-redshift tracers of structure.

The templates are built for a single source distance, whereas the forest pixels span a range of distances; the first-order correction for this, which modifies the geometric factor $G_{ab}$ through a derivative map that accompanies every template, is described in Section~\ref{sec:sourcedist} after we introduce the template generation formalism.

There is one particular subtlety that is worth pointing out. If templates are noiseless, the equations apply as written. If templates are noisy, the estimator is still valid, but in general needs an amplitude calibration. For a given template $\alpha^T$ and a true displacement $\alpha^{\rm truth}$, the expectation value of the estimator goes as (assume a single template for simplicity) $\left<\hat{A}\right> = \left<\alpha^T \alpha^{\rm truth}\right>/\left<\alpha^T \alpha^T \right>$. For an exact Wiener filter with the correct signal and noise covariances this ratio is unity (Appendix~\ref{app:wiener}). This motivates the the Wiener construction below, whose normalization is checked empirically against CMB lensing in Section \ref{sec:cmbval}.

\section{Templates from DESI and BOSS}
\label{sec:templates}
\subsection{Data Used}
\subsubsection{Low-redshift tracers}
\label{sec:lensdata}

\begin{figure*}
\centering
%% To reproduce: python scripts/report_figures.py (LyaLenser repository; reads $LYALENSER_DATA/lowz_v4/summary.json and computes the Limber spectra of lyalenser/lowz.py) -> report/figures/kernel_slices.pdf
\includegraphics[width=\linewidth]{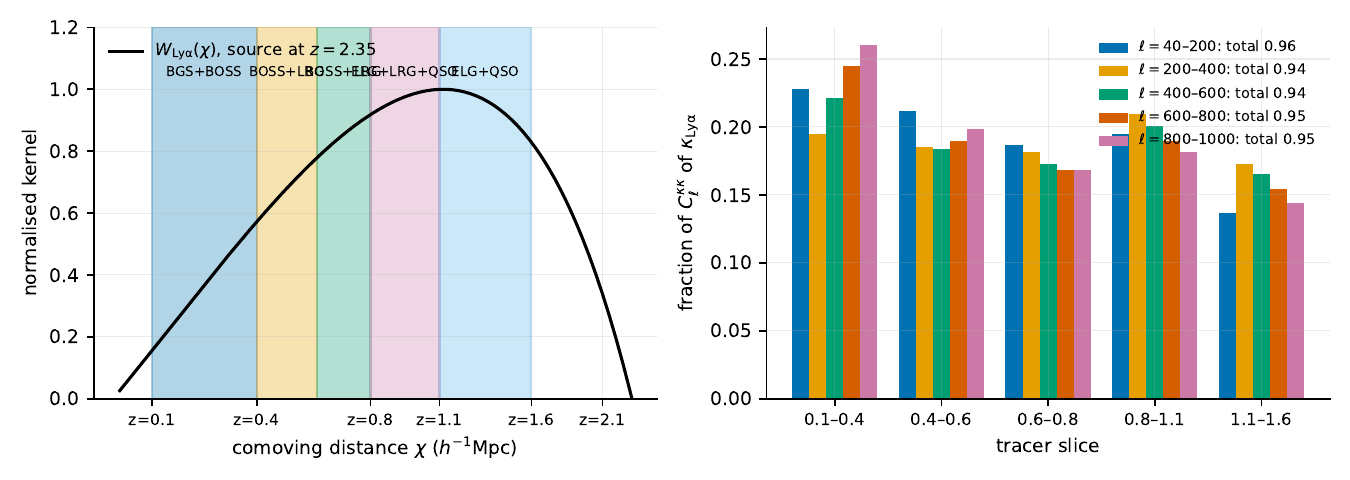}
\caption{Left: the lensing kernel of a source at $z_{\rm eff}=2.35$, the weighted mean pixel redshift of the forest sample, with the five tracer slices shaded and their tracers named. Right: the fraction of the convergence power $C_\ell^{\kappa\kappa}$ of $\kl$ contributed by each slice, averaged over each of the five science bands (Limber, non-linear matter power); the five slices together hold 94--96 per cent in every band, the lowest slice alone 20--26 per cent.}
\label{fig:kernel}
\end{figure*}

Three catalogues provide the lenses. (i) The DESI DR1 large-scale-structure clustering catalogues v1.5 \citep{2405.16593}: LRG ($0.4<z<1.1$), ELG ($0.8<z<1.6$), QSO ($0.8<z<1.6$ here) and the magnitude-limited bright-galaxy sample BGS ($0.1<z<0.4$), with their completeness, redshift-failure and imaging-systematics weights, and two random catalogues per Galactic cap. (ii) The BOSS DR12 combined LOWZ+CMASS galaxy sample \citep{2016MNRAS.455.1553R}, $0.1<z<0.8$, with its systematic, close-pair and redshift-failure weights and one random catalogue per cap; BOSS covers sky that DESI DR1 does not (and vice versa). They are split into five redshift slices, $0.1$--$0.4$ (BGS, BOSS), $0.4$--$0.6$ (LRG, BOSS), $0.6$--$0.8$ (LRG, BOSS), $0.8$--$1.1$ (LRG, ELG, QSO) and $1.1$--$1.6$ (ELG, QSO). There is approximately $400\Mpch$ buffer between our highest redshift slice and beginning of the lensing plane sources to suppress direct density correlations. Figure~\ref{fig:kernel} shows the slices on the lensing kernel of the forest and the share of the convergence power each of them carries.

\subsubsection{CMB lensing}
The ACT DR6 baseline convergence map \citep{2304.05203,2304.05202} and the Planck PR4 convergence map \citep[the minimum-variance reconstruction of][rotated from Galactic to equatorial coordinates]{2022JCAP...09..039C} are used to validate the tracer biases and the deflection templates. They are not used in the lensing measurement itself.

\subsection{Building the templates}
Low-redshift galaxies trace the matter density that lenses the \lya forest and are used to generate the templates. On large scales, galaxy number density fluctuations $\delta_g$ directly trace the matter density fluctuations $\delta_m$ with a linear bias
\begin{equation}
    \delta_g = b\delta_m + n,
\end{equation}
where $b$ is the bias parameter and $n$ is noise dominated by Poisson noise. On smaller scales the relation is more complicated and we do not attempt to model it.
Our approach is to use the auto-correlation of each tracer to measure its bias (against a fixed cosmological model) and then to use this bias to generate templates that are appropriately normalised with respect to the \lya forest lensing plane. We use relatively thin redshift slices to take into account the evolution of bias with redshift.

Because the DESI data are affected by fibre collisions and other vagaries, and because the signal-to-noise of the auto-correlations is not a limiting factor, we split every galaxy sample into two random halves and cross-correlate them to measure the power spectrum -- this ensures that the shot noise cancels. We use NaMaster \citep{1809.09603} to calculate the spectra: the bandpowers (width $\Delta\ell=40$) are decoupled from the binary survey mask, the theory is pushed through the same bandpower windows, and the bandpower covariance is the Gaussian estimate with mode coupling included.

To measure the scale dependence, we split the template deflection into five multipole ranges, $40\le \ell<200$, $200$--$400$, $400$--$600$, $600$--$800$ and $800$--$1000$ (cosine-tapered windows; the templates carry $\ell\le1000$), to which we add the same number of ``curl'' maps (the template deflection rotated by 90 degrees, as a systematic check) and one last band to mop up the power outside the science windows in the template map.

\begin{figure*}
\centering
%% To reproduce: python scripts/report_figures.py (LyaLenser repository; reads $LYALENSER_DATA/lowz_v3/summary.json, whose bias fits are those of the fiducial lowz_v4 templates) -> report/figures/tracer_spectra.pdf, copied here.
\includegraphics[width=\linewidth]{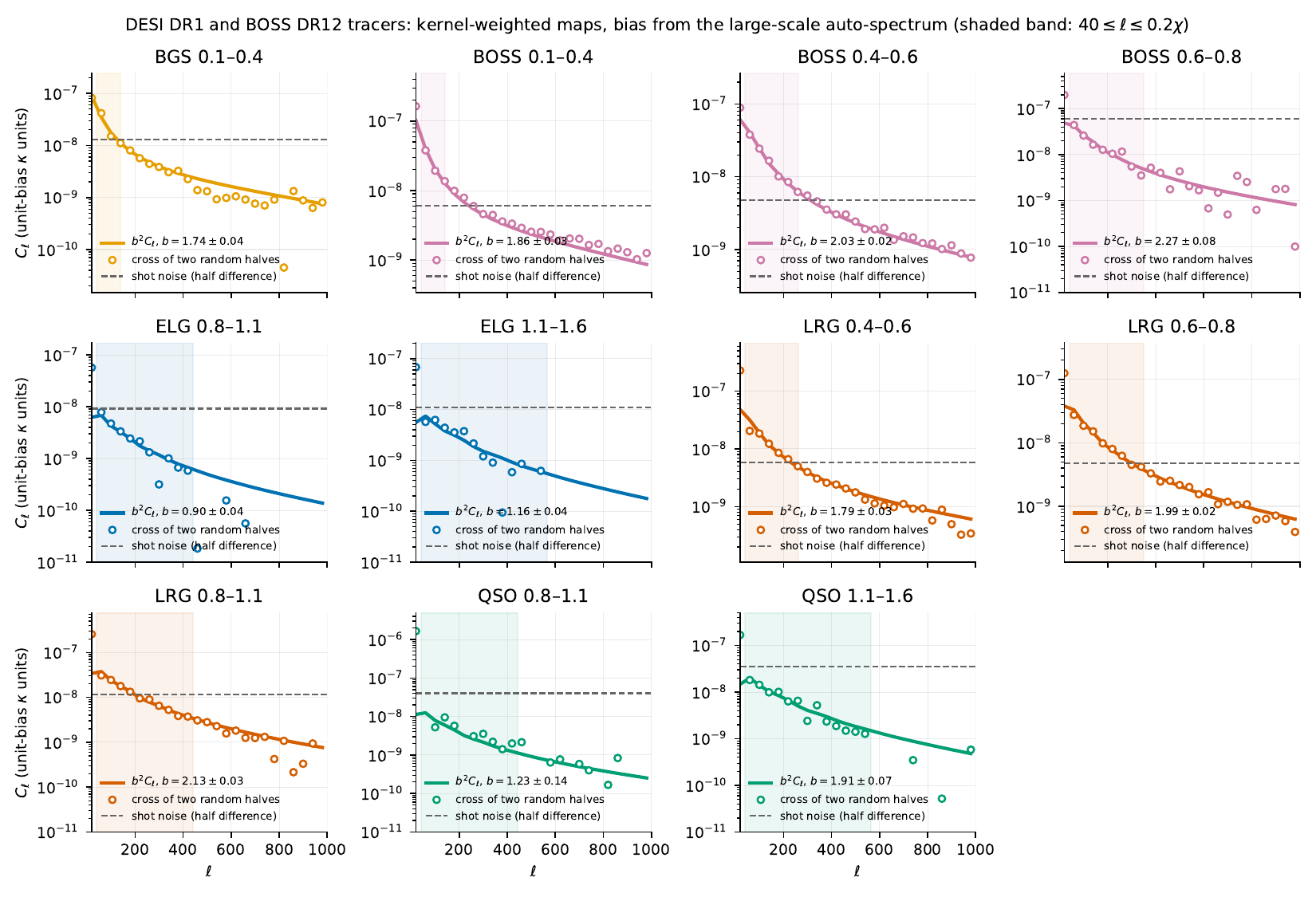}
\caption{Angular power spectra of the eleven tracer maps in convergence units for unit bias: the NaMaster cross-spectrum of the two random halves of each catalogue (points, free of shot noise), the fitted $b^2C_\ell$ of Equation~\ref{eq:limber} pushed through the bandpower windows (solid line) and the shot noise measured from the half-difference map (dashed). The shaded band is the fit range $40\le\ell\le0.2\,\chi(z_{\rm mid})$.}
\label{fig:spectra}
\end{figure*}

In detail, the per-tracer templates are built as follows. The tracers are divided into redshift slices $[z_1,z_2]$. For a tracer with objects at $(\vth_i,\chi_i)$, catalogue weights $w_i$ and mean density $\bar n(\chi)$ per steradian per unit comoving distance, the kernel-weighted map
\begin{equation}
m(\vth)=\sum_{i\in{\rm pixel}}\frac{w_i\,W(\chi_i)}{\bar n(\chi_i)\,\Omega_{\rm pix}}-\text{(randoms, scaled)},
\label{eq:map}
\end{equation}
with the lensing kernel for a source at the effective forest distance $\chi_s=\chi(z_{{\rm Ly}\alpha})$,
\begin{equation}
W(\chi)=\frac{3}{2}\Omega_m\frac{H_0^2}{c^2}(1+z)\,\chi\,\frac{\chi_s-\chi}{\chi_s},
\label{eq:kernel}
\end{equation}
estimates bias times the contribution of the slice to the forest convergence, $\langle m\rangle=b\int_{z_1}^{z_2}W\,\delta_m\,d\chi$, with $z_{{\rm Ly}\alpha}=2.35$ the weighted mean pixel redshift of the forest sample (Section~\ref{sec:data}). The randoms define the footprint (HEALPix pixels at $\Nside=512$ holding at least half the mean random count, which excludes partially covered rim pixels), the completeness (the random density smoothed by $1^\circ$, by which the map is divided) and the mean-density subtraction. The linear bias of every (tracer, slice) is measured from the map itself: the cross-spectrum of the two random halves is fitted, on large scales only ($40\le\ell\le \ell_{\rm max}$ with $\ell_{\rm max}$ set by $k_{\rm max} = 0.2\,h\,{\rm Mpc}^{-1}$), against the non-linear matter power spectrum computed with HMcode-2020 \citep{2009.01858} (in Limber approximation)
\begin{equation}
C_\ell=b^2\int_{z_1}^{z_2}\frac{W(\chi)^2}{\chi^2}\,P\!\left(k=\frac{\ell+1/2}{\chi},z\right)d\chi
\label{eq:limber}
\end{equation}
times the pixel window. Dividing the map by $\hat b$ gives the tracer's estimate $\hat\kappa_k$ of the slice's convergence, and the shot noise $N_k$ of that estimate is measured as the white level of the spectrum of the difference of the two half-maps. Figure~\ref{fig:spectra} shows the measured spectra with the fits and the shot-noise levels.

\begin{figure}
\centering
%% To reproduce: python scripts/report_figures.py (LyaLenser repository; reads $LYALENSER_DATA/lowz_v3/summary.json and results/cmb_bias_check{,_planck}.json) -> report/figures/biases.pdf, copied here.
\includegraphics[width=\linewidth]{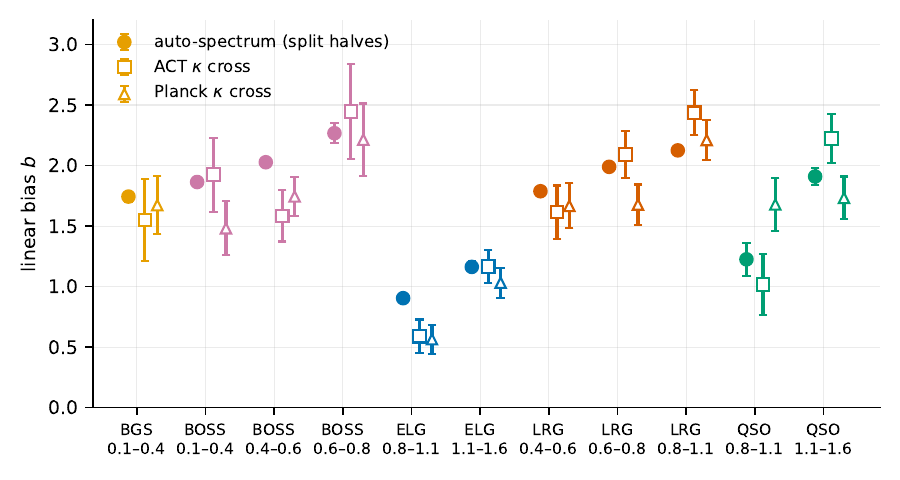}
\caption{The linear bias of the eleven tracer samples from the auto-spectrum of Figure~\ref{fig:spectra} (filled circles), from the cross-spectrum with the ACT convergence map (squares) and with the Planck map (triangles).}
\label{fig:bias}
\end{figure}

As a sanity check we also measure the biases from the cross-spectra of the tracer maps with the ACT and Planck convergence maps, $\langle m\kc\rangle_\ell=b\int W\,W_{\rm CMB}\,P\,d\chi/\chi^2$, which need no shot-noise model. Figure~\ref{fig:bias} compares the three estimates: they agree within their errors for every sample, with the cross-correlation biases lower by 4 per cent (ACT) and 7 per cent (Planck) on average, the same deficit seen in the template validation below. The maps are normalised with the more precise auto-spectrum biases.

At this point we need to deal with the fact that the coverage is not uniform: a pixel may be covered by only two of the three tracers of a slice, or by one. We therefore partition the footprint of a slice into \emph{coverage classes}, the distinct subsets of tracers covering a pixel, cut each tracer map to each class in pixel space, transform to harmonic space and combine with per-$\ell$ Wiener weights,
\begin{equation}
\hat\kappa_{\rm slice}(\ell)=\sum_k w_k(\ell)\,\hat\kappa_k(\ell)
\end{equation}
with $\bm w_\ell=\mathsf C_\ell^{-1}\bm s_\ell$ computed per class, where $\mathsf C_{kl}(\ell)$ is the covariance of the maps for the tracers $k$ and $l$ present in the class and $s_k(\ell)=S_\ell p_\ell$ is their expected cross-spectrum with the true convergence ($S_\ell$ the slice's convergence spectrum, $p_\ell$ the pixel window). This construction approximates the Wiener estimate using multipole-dependent weights. The tracers of a slice trace the same density field and sometimes even the same objects (e.g. where BOSS and DESI overlap), so their maps are strongly correlated and this inverse covariance weighting accounts for their measured correlations. We measure $\mathsf C_{kl}$ directly, as the full set of NaMaster auto- and cross-spectra of the unit-bias maps on the footprint common to the tracers of the slice, binned and interpolated to every multipole.

\begin{figure}
\centering
%% To reproduce: python scripts/plot_templates.py --lowz $LYALENSER_DATA/lowz_v4 --combined-only (LyaLenser repository) -> figures/templates_map_combined.pdf
\includegraphics[width=\linewidth]{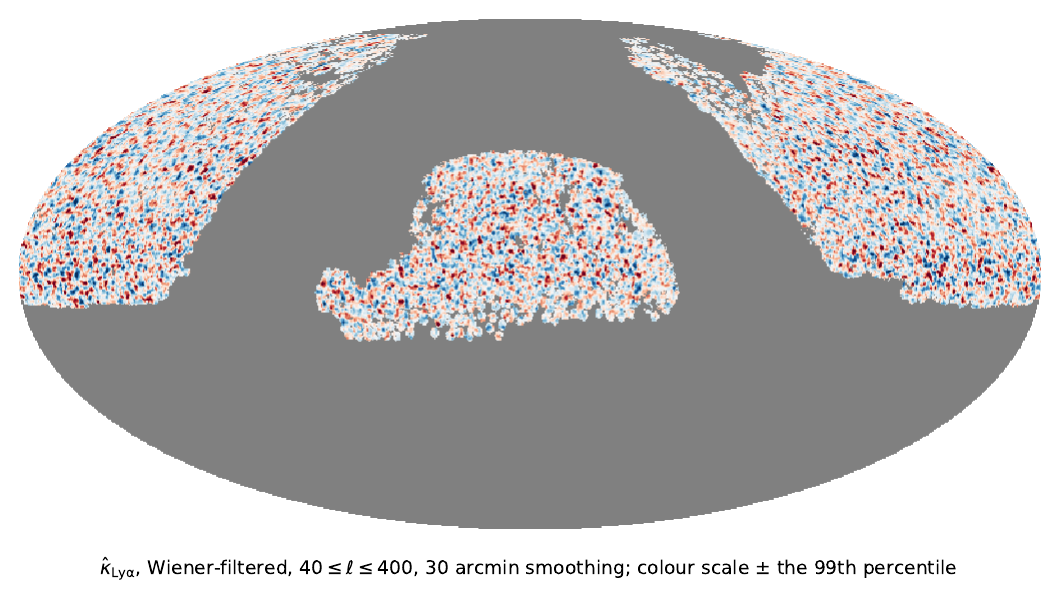}
\caption{The combined convergence template, the sum of the five Wiener-filtered slice maps, in equatorial Mollweide projection, filtered to $40\le\ell\le400$ and smoothed by 30 arcmin for display; the colour scale is $\pm$ the 99th percentile. The footprint is the union of the DESI DR1 and BOSS DR12 tracer footprints.}
\label{fig:tmaps}
\end{figure}

Each slice thus yields one Wiener-filtered convergence map, and the sum of the five is our best estimate of the $\kappa$ field at the \lya forest plane, the combined template shown in Figure~\ref{fig:tmaps}. The deflection of each slice map, split into the five science bands, their curl partners and the junk band, gives 11 template components per slice and 55 in total, which are the components of the joint fit of Section~\ref{sec:results}.

\subsection{Source-distance dependence of the templates}
\label{sec:sourcedist}
%% Numbers in this subsection: results/source_distance_expansion.json (python scripts/source_distance_expansion.py --run $LYALENSER_DATA/stageb/dr1_lowz_v8 --lowz $LYALENSER_DATA/lowz_v5 --summary results/dr1_lowz_v8.json --previous results/dr1_lowz_v7d.json) and results/joint_fit_v5.json, joint_fit_v5_noderiv.json
The templates are evaluated for a single source distance $\chi_s=\chi(z_{{\rm Ly}\alpha})$, the weighted mean pixel distance of the forest sample, whereas the two pixels of a pair sit at $\chi_p$ and $\chi_q$ and are lensed with the efficiency of their own distance. We include this to first order in the distance offset. The kernel of Equation~\ref{eq:kernel} depends on the source distance through
\begin{equation}
\frac{\partial W}{\partial\chi_s}=\frac{3}{2}\Omega_m\frac{H_0^2}{c^2}(1+z)\,\frac{\chi^2}{\chi_s^2},
\label{eq:dkernel}
\end{equation}
so alongside every template we build its \emph{derivative map}: the same tracers weighted by $\partial W/\partial\chi_s$ in place of $W$ in Equation~\ref{eq:map}, divided by the same bias and combined with the same Wiener weights, which is exact because the construction is linear in the map. Its deflection $\bm\alpha'^T=\partial\bm\alpha^T/\partial\chi_s$ is evaluated at the sightlines in the same multipole bands, and the deflection of a pixel at distance $\chi$ is
\begin{equation}
\bm\alpha(\vth,\chi)\simeq\bm\alpha^T(\vth)+(\chi-\chi_s)\,\bm\alpha'^T(\vth).
\label{eq:dalpha}
\end{equation}
Writing $\chi_{p,q}=\chi\pm\Delta\chi/2$, with $\chi$ the pair mean distance, the geometric factor of Equation~\ref{eq:pair} becomes
\begin{multline}
G_{ab}=\chi\,\hat\vth_{ab}\cdot\Bigl\{(\bm\alpha_a-\bm\alpha_b)+(\chi-\chi_s)(\bm\alpha'_a-\bm\alpha'_b)\\
+\tfrac12\Delta\chi\,(\bm\alpha'_a+\bm\alpha'_b)\Bigr\},
\label{eq:g1pair}
\end{multline}
so the derivative map enters through both its difference and its sum over the pair. The sums of Equation~\ref{eq:qe} are accumulated together with the moments of $\chi-\chi_s$ and $\Delta\chi$ that these terms require, and the correction is carried through as any other weighting (i.e. both the raw estimator and the response matrix). The size of the corrections grows with redshift.  Over the weighted rms of the pixel distances about $\chi_s$, $210\Mpch$, the deflection of the lowest slice varies by 1 per cent rms and that of the highest by 14 per cent. Despite this, the effect still mostly averages out leaving the result essentially unchanged even if the effect is completely neglected (see Figure \ref{fig:robust} below). One place where the term matters is the redshift split of Section~\ref{sec:amplitude}, where the pixels of a sub-range sit systematically in front of or behind $\chi_s$, than for the full sample.

\subsection{Validating templates with CMB lensing cross-correlation}
\label{sec:cmbval}
To validate the deflection templates we cross-correlate them with CMB lensing convergence maps, which are well established and calibrated: the ACT DR6 baseline map and the Planck PR4 map (Section~\ref{sec:lensdata}). We test the product the estimator consumes, the band-filtered deflection $\bm\alpha^T$ of the Wiener-filtered template, rather than the tracer maps: the deflection is built on the sphere for every science band and for the whole window, and cross-correlated as a spin-1 field with the convergence maps using NaMaster (bandpowers of width 40, decoupled from the template mask and from the CMB mask, the latter squared on the ACT side as the release prescribes; errors from the NaMaster Gaussian covariance with measured and smoothed auto and cross spectra). The E-mode of the deflection carries the signal; for a pure gradient field the B-mode is mask leakage only and serves as a null.

The expectation is not the $\kl\times\kc$ spectrum, because the template estimates only the part of $\kl$ traced by objects at $0.1<z<1.6$ and is Wiener-suppressed where those tracers are noisy.  Moreover, different tracers have different overlaps with the Planck and ACT maps. The prediction for the cross-spectrum of the template convergence $T$ with the CMB convergence is therefore

\begin{equation}
C_\ell^{T\times\kappa_{\rm CMB}} =\sum_s\;\sum_c f_c \sum_{k\in c} W^c_k(\ell)\;C_\ell^{(s,\rm CMB)}\,p_\ell,
\label{eq:pred}
\end{equation}
where $s$ runs over the redshift slices, $c$ over the coverage classes of the slice, $k$ over the tracers of the class, $f_c$ is the fraction of the template--CMB overlap occupied by class $c$, $W^c_k$ are the Wiener weights of the class, $C_\ell^{(s,\rm CMB)}$ is the Limber cross-spectrum of the slice's contribution to $\kl$ with $\kc$, and $p_\ell$ is the pixel window. The deflection E-mode of a band follows by the factor $2/\sqrt{\ell(\ell+1)}$ and the band window. 
The amplitude $A_L$ of the measured cross-spectrum relative to this prediction is expected to be one if the biases, the covariance and the Wiener weights are right, independently of the fact that the sources are at different redshifts. These corrections are far from trivial: the ratio of the prediction to the full $\kl\times\kc$ spectrum is 0.07--0.43 depending on the band. Note that even though we cross-correlate galaxies with CMB to cross-check the bias values, the final biases used were auto-correlation derived. Moreover, we cross-correlate on scales on which the field becomes quasi non-liner and and there is no a-priori reasons that these tests should pass.

\begin{figure*}
\centering
%% To reproduce: python scripts/deflection_cmb_check.py --lowz $LYALENSER_DATA/lowz_v4 --bands 40 200 400 600 800 1000 --out results/deflection_cmb_check_v4.json; python scripts/plot_template_cmb_cross.py --tag v4 (LyaLenser repository) -> figures/template_cmb_cross.pdf
\includegraphics[width=\linewidth]{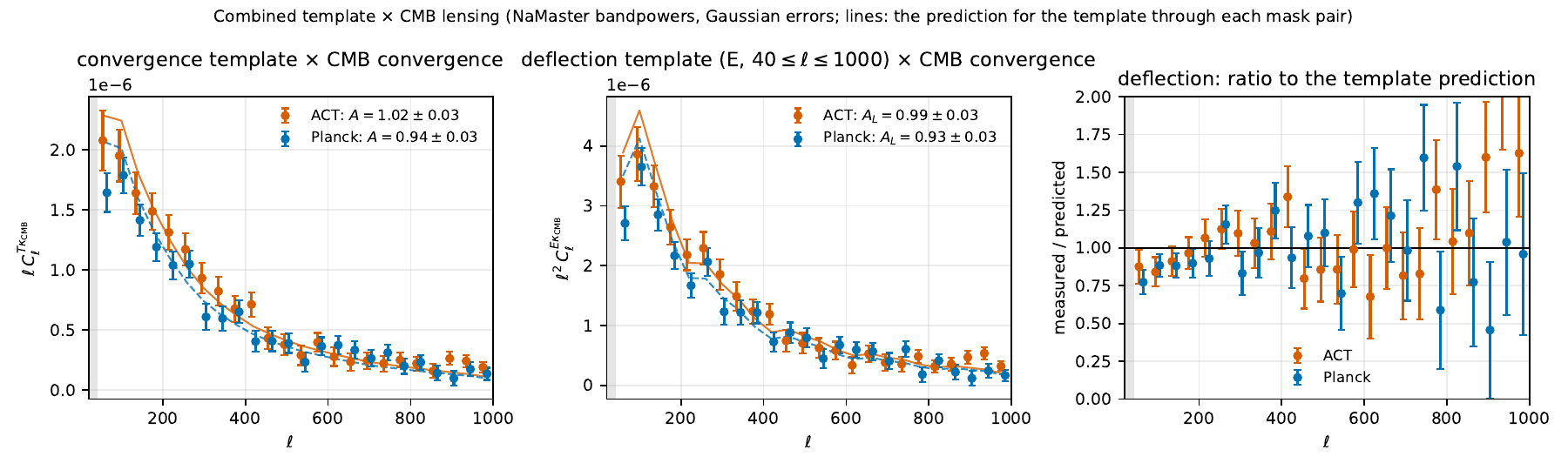}
\caption{The combined template against the ACT (orange) and Planck (blue) convergence maps, NaMaster bandpowers with Gaussian errors on the full overlap with each map and the prediction of Equation~\ref{eq:pred} through each mask pair (lines). Left: the convergence template. Middle: the E-mode of the band-filtered deflection template over the science window $40\le \ell\le1000$, the field the estimator consumes. Right: the ratio of the deflection cross-spectrum to its prediction.}
\label{fig:tcc}
\end{figure*}

Figure~\ref{fig:tcc} shows the results of correlation against the convergence map and the deflection map. The non-smooth structure in the deflection map is associated with edges in the filtering ranges. In total we find  $A_L=0.99\pm0.04$ against ACT and $0.93\pm0.03$  against Planck for the combined deflection template over the science window $40\le\ell\le1000$. Visually, there is some evidence that the templates are low at the largest scales and the measured Planck value is about $2\sigma$ low. Given the expected error on \lya lensing, however, these templates are plenty accurate enough.

\section{\lya forest data and response}
\label{sec:data}

\subsection{The \lya forest and quasar data}
We use the \lya forest flux fluctuations of the DESI Data Release 1 \citep{2503.14745}, extracted with the standard DESI pipeline \texttt{picca} \citep{2306.06312} on a 0.8\,\AA\ grid, with the DLA and BAL masks applied by the release. The catalogue provides two rest-frame windows and we use both: region A, $1040$--$1205$\,\AA, the \lya forest proper, and region B, $920$--$1020$\,\AA, blueward of the Lyman-$\beta$ emission line. A region-B pixel at observed wavelength $\lambda$ carries \lya absorption at $z=\lambda/1215.67-1$, the field we want, and simultaneously Ly$\beta$ absorption from $z=\lambda/1025.72-1$.  A pixel pair inside $30\Mpch$ that joins an A and a B pixel therefore measures the \lya--\lya correlation alone, while a B--B pair also carries Ly$\beta$--Ly$\beta$ correlation at a common redshift; we use A$\times$A and A$\times$B pairs and drop B$\times$B. Region B is treated as an extension of the same sightline, each segment keeping its own continuum-fit block. We keep every pixel of the DR1 grid below $z=3$: the grid starts at 3600\,\AA, i.e.\ $z=1.96$, in both windows, so the sample is $1.96\le z\le3.0$; per region, forests with at least 50 such pixels are kept. This gives 427\,888 sightlines with $2.8\times10^8$ pixels (19.5 per cent in region B) over $11\,000$\,deg$^2$, with an average sightline density of $38.7\,$deg$^{-2}$ over the full selected redshift range. The weighted mean pixel redshift, $z_{\rm eff}=2.348$ (unweighted mean 2.342, median 2.292), is adopted as the effective forest redshift and as the source plane of the templates. For the cross-correlation we use the DR1 quasar catalogue at $1.9<z<3.1$, one entry per target (583\,148 quasars, 377\,715 of them the quasars of forests in the sample), which gives $9.5\times10^6$ quasar--sightline pairs within $30\Mpch$. Pixel weights are the \texttt{picca} inverse-variance weights $w=1/(\eta\sigma_N^2+\sigma^2_{\rm LSS}(\lambda))$, which include the intrinsic variance of the flux field.

\subsection{The forest-forest response}
\label{sec:kernel}

To evaluate the response kernel $g=G\,\xi_F'$ we first need $\xi_F$ itself. We measure $\xi_F(\rperp,\rpar)$ from the same forests in $1\Mpch$ cells over $\rperp,\rpar<30\Mpch$, and, because of the strong evolution of the \lya forest flux correlation function, the cells are further split by the mean redshift of the pair, in seven bins between $z=1.96$ and $3.0$ (the pair weight falls by a factor of seven from the first to the last bin), recording the pair-weighted mean redshift of every cell. 

We then fit these measurements using a simple phenomenological model.  All we need is a smooth, analytically differentiable model that fits the data well enough to give a self-consistent response. We use
\begin{multline}
\xi_F(\rperp,\rpar,z)=b_F(z)^2\,\xi_{\rm ref}\big(\rperp,\rpar;\beta_F(z)\big)\\
+\left(\frac{1+z}{1+z_{\rm ref}}\right)^{\gamma_s} \sigma(r,z)\,S(\rperp,\rpar)+N(\rperp)
\label{eq:xifit}
\end{multline}
where
\begin{eqnarray}
    P_F(\bm k)=(1+\beta_F\mu^2)^2P(k,z_{\rm ref}) \\
    b_F(z) = b_F \,\frac{D(z)}{D(z_{\rm ref})} \left(\frac{1+z}{1+z_{\rm ref}}\right)^{\gamma_b}\\
    \beta_F(z) = \beta_F\left(\frac{1+z}{1+z_{\rm ref}}\right)^{\gamma_\beta}
\end{eqnarray}
Here $\xi_{\rm ref}$ is the linear model given by $P_F(\bm k)$ pushed through the DESI pixel and resolution windows and the per-forest continuum operator, which takes into account that the continuum fitting removes the slowly varying modes of each forest \citep{1702.00176,2007.08995}. 
The linear model alone is insufficiently accurate  so we add the additional pieces: $S$ is a product of cubic B-splines in $\rperp$ and $\rpar^2$ carried on a strictly positive envelope $\sigma(r=(\rperp^2+\rpar^2)^{1/2})$ (the monopole of $\xi_{\rm ref}$), which removes the dynamic range of $\xi_F$ from the spline coefficients; and $N$ is a one-dimensional spline confined to $\rpar<1\Mpch$ that absorbs the excess systematic correlations that appear at the same wavelength.

Importantly, since we associate $N$ with systematics rather than real signal, we ignore it when taking the $\rperp$ derivative; this has per-cent-level effects on the results (see Section~\ref{sec:injected1}).

\begin{figure*}
\centering
%% To reproduce: python scripts/paper_xi_figures.py (LyaLenser repository; reads $LYALENSER_DATA/stageb/dr1_lowz_v8/xi.h5 and dr1_qso_v2/xi_qf.h5, the tables of dr1_lowz_v7d and dr1_qso_v1d) -> figures/xi_ff_fit.pdf
\includegraphics[width=\linewidth]{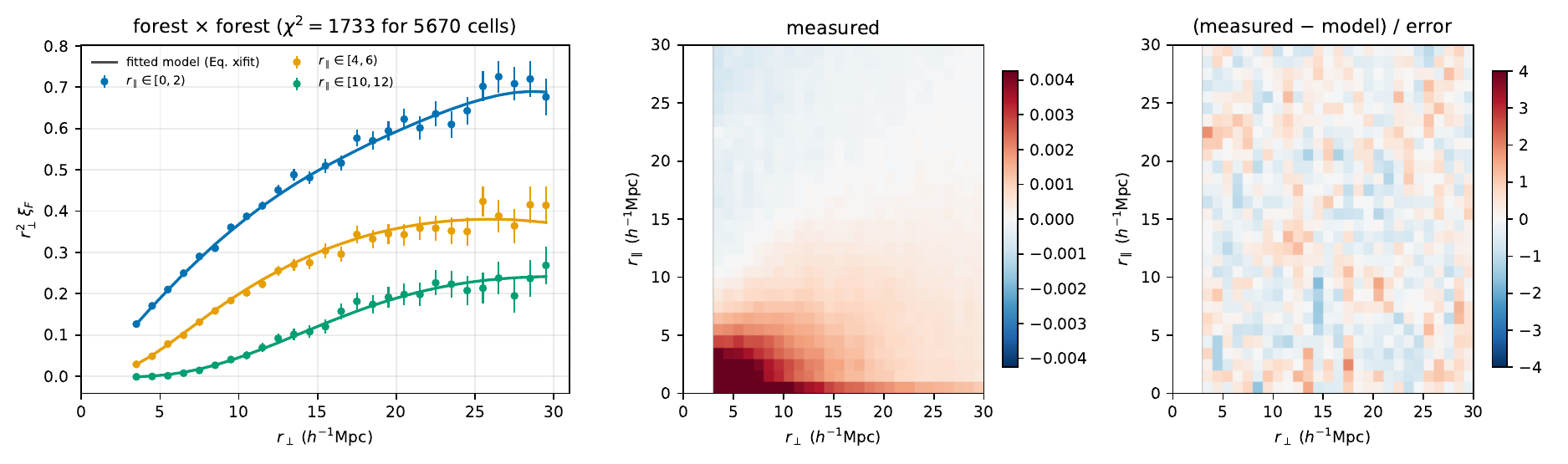}
\caption{The forest auto-correlation that sets the forest--forest response kernel. Left: the measured $\rperp^2\xi_F$ in three strips of $\rpar$ (points, pair-weight errors), collapsed over the seven redshift bins with the pair weights, and the fitted model of Equation~\ref{eq:xifit} collapsed the same way (lines). Middle: the measured $\xi_F$ on the $(\rperp,\rpar)$ plane over the fit range $3\le\rperp<30\Mpch$. Right: the residual of the fit in units of the per-cell error.}
\label{fig:xiff}
\end{figure*}

A single evolving model is then fitted to all $7\times810$ cells with the accumulated pair weights: the five non-linear parameters $(b_F^2,\beta_F,\gamma_b,\gamma_\beta,\gamma_S)$ are fitted by least squares while the linear spline coefficients are solved exactly at every step. The fit gives $\gamma_b=3.8$ and $\beta_F\propto(1+z)^{-2.1}$ with $\beta_F=0.59$ at $z=2.4$, and $\chi^2=1733$ for 5643 degrees of freedom with the pair-weight errors. Figure~\ref{fig:xiff} compares the fitted model with the measured cells, collapsed over redshift with the pair weights (the model collapsed the same way): the residuals show no structure at the level of the per-cell error.

\emph{The continuum projection.} The correlation the pixels see is $P_aCP_b^\mathsf{T}$ with $P_a$ the mean-and-slope removal of the continuum fit on forest $a$; it is the largest single distortion of the model. Evaluating it for one representative forest spanning the whole slab with uniform weights over-predicts $\xi_F$ by 5--20 per cent, growing with $\rperp$; we therefore use the pair-weight average of $P_aCP_b^\mathsf{T}$ over sampled pairs of real DR1 forests, with the measured pixel weights and with the projector running over every pixel in the data (not only the pixels inside $1.96<z<3.0$). 

\emph{How much freedom to give the correction.} There is an important subtlety in this analysis: the kernel is the derivative of a correlation function fitted to noisy cells, and random errors in it can affect the normalization as well as the variance. Under simplifying assumptions they attenuate the amplitude, because they enter the response quadratically but the signal only linearly (Appendix~\ref{app:kernel_error}). More free parameters remove misfit but add fitted noise. We adopt a bicubic surface in $(\rperp,\rpar^2)$ with 22 fitted parameters in all -- we later test for this choice and find no sensitivity (see Figure \ref{fig:robust})

\subsection{The forest-quasar response}
\begin{figure*}
\centering
%% To reproduce: python scripts/paper_xi_figures.py (LyaLenser repository) -> figures/xi_qf_fit.pdf
\includegraphics[width=\linewidth]{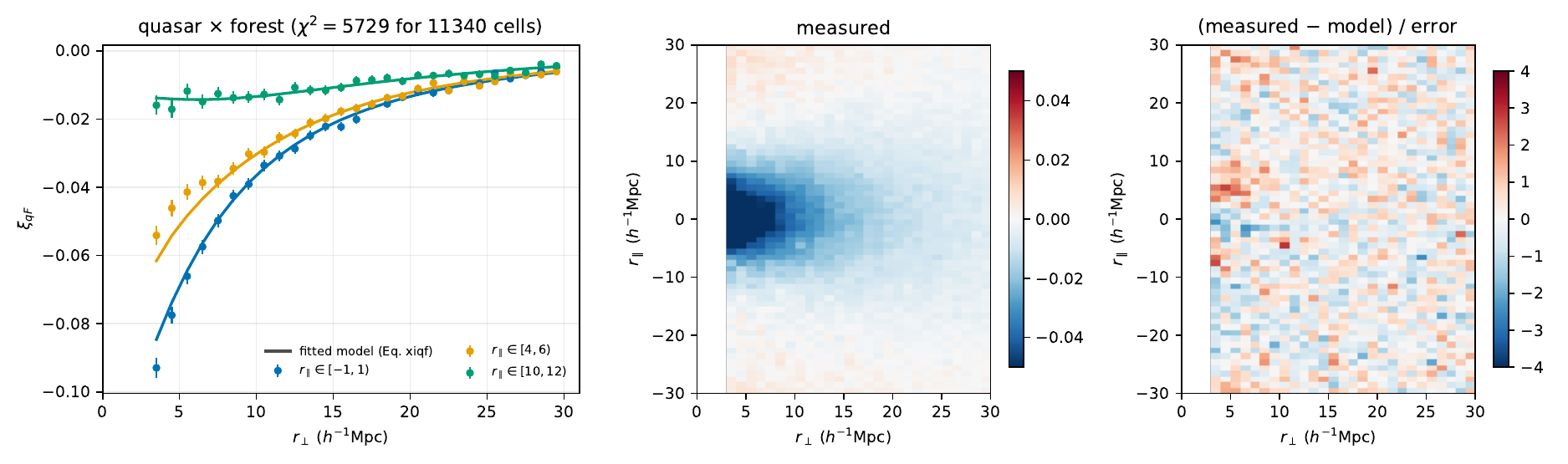}
\caption{The same for the quasar--forest cross-correlation, with $\rpar$ signed (pixel minus quasar) and the model of Equation~\ref{eq:xiqf}. The profiles in the left panel are for $\rpar$ strips centred on $0$, $5$ and $11\Mpch$; the asymmetry of the middle panel about $\rpar=0$ is the fitted quasar redshift offset.}
\label{fig:xiqf}
\end{figure*}

The approach with $\xi_{qF}$ is analogous to that of $\xi_{F}$.  $\xi_{qF}$ is measured in $1\Mpch$ cells over $r_\perp<40$, $|r_\parallel|<40\Mpch$ (signed, pixel minus quasar) and the same seven bins of pair mean redshift, from $3.6\times10^7$ quasar--sightline pairs, and fitted over $3\le\rperp<30$, $|\rpar|<30\Mpch$ ($7\times1620$ cells) with an independent model of the same construction:
\begin{multline}
\xi_{qF}(\rperp,\rpar,z)=\\
-b_F(z)b_q(z)\,\xi_{\rm ref}\big(\rperp,\rpar-\Delta \rpar;\beta_F(z);\beta_q(z)\big)\otimes\mathcal N(\sigma_\parallel)\\
+\left(\frac{1+z}{1+z_{\rm ref}}\right)^{\gamma_s} \sigma(r,z)\,S(\rperp,\rpar)
\label{eq:xiqf}
\end{multline}
with $b_q$ similarly evolving with redshift with the power law index $\gamma_q$, but with quasar redshift-space distortion given by the standard  $\beta_q = f(z)/b_q(z)$. Note that we choose to define both $b_F$ and $b_q$ as positive quantities and therefore need to add an extra minus sign to this correlation function. We also fit for $\Delta \rpar$ to account for the systematic quasar redshift offset \citep{2007.08995,2404.03001} and convolve the resulting function with a Gaussian of width $\sigma_\parallel$ to account for the quasar redshift errors. There is no same-wavelength term for quasars, and the spline coefficients are fitted separately for the two correlations. The forest parameters are held at the values of a base-only auto-correlation fit, without the spline correction. We first fit $(b_q,\gamma_q,\Delta\rpar,\sigma_\parallel)$ without a cross-correlation spline, obtaining $b_q=3.20$ at $z=2.4$, $\gamma_q=1.7$, $\Delta\rpar=0.44\Mpch$ and $\sigma_\parallel=4.9\Mpch$. Holding these parameters fixed, we then fit $\gamma_S$ and the 20 spline coefficients, obtaining $\chi^2=5729$ for 11\,319 nominal degrees of freedom conditional on the base fit. Figure~\ref{fig:xiqf} shows the fit; the quasar redshift offset and smoothing are visible as the shift and rounding of the peak of the profiles about $\rpar=0$.

\section{Results}
\label{sec:results}

\subsection{Validation using injected signal}
\label{sec:injection}
We test the geometric response by shifting every position (the sightlines, and for the cross-correlation also the quasars) by $-A\,\bm\alpha_{\rm inj}(\vth)$, rebuilding the pairs with the production selection, including the exclusion of B$\times$B pairs, and re-running the estimator on the same pixels. We take $\bm\alpha_{\rm inj}$ to be the science-band deflection of the combined template itself and $A=\pm0.25,\pm0.5$. These checks use the combined template's eleven components, with a common science amplitude and free curl and junk amplitudes. Therefore they test the estimator's response to that particular template, not its calibration against an independent true lensing field or the full joint slice fit. We quote the slope $[\hat A(+a)-\hat A(-a)]/2a$, which cancels contributions unchanged by the shift, and calculate its jackknife error from matched leave-one-region-out estimates at the different shifts.

\subsubsection{Noiseless injection}
\label{sec:injected1}
%% Numbers of this and the next subsection: results/injection_v5.json (python scripts/injection_dr1.py --auto $LYALENSER_DATA/stageb/dr1_lowz_v8 --cross $LYALENSER_DATA/stageb/dr1_qso_v2 --lowz $LYALENSER_DATA/lowz_v5 --bands 40 200 400 600 800 1000 --tag v5) and results/injection_v5_nosw.json (--remove-same-wavelength --statistics auto --modes expectation)
In the first version of the test the product $\dF(p)\dF(q)$ (or $\dF(p)$ for the cross-correlation) is replaced by the fitted correlation at the true, unshifted separation, while the response kernel, the cuts and the mean field use the shifted geometry. This is a noise-free test of the signs, local metric, pair selection and band-basis representation of the injected deflection. The recovered slopes are $1.025$ (forest$\times$forest) and $0.998$ (quasar$\times$forest). The injection displaces the whole fitted table, including the same-wavelength term $N(\rperp)$ whose derivative is deliberately omitted from the lensing kernel. Removing this term from the injected correlation gives an auto slope of $0.991$ at $|A|=0.25$. This demonstrates the full estimator is at most biased at percent level, negligible compared to our sensitivity.

\subsubsection{Actual injection}
In the second version the shifts are applied to the real data. The recovered slopes are $1.060\pm0.040$ (auto) and $0.999\pm0.048$ (cross), with errors from the paired jackknife. Both are compatible with unity, and the auto slope is also consistent with the full-table noiseless response. The mean curl amplitudes have slopes consistent with zero.

\subsection{Scale sensitivity}

\begin{figure*}
\centering
%% To reproduce: python scripts/scale_sensitivity.py --auto $LYALENSER_DATA/stageb/dr1_lowz_v8 --cross $LYALENSER_DATA/stageb/dr1_qso_v2 --lowz $LYALENSER_DATA/lowz_v5 --bands 40 200 400 600 800 1000 --tag v5 (LyaLenser repository) -> figures/scale_sensitivity.pdf, numbers in results/scale_sensitivity_v5.json
\includegraphics[width=\linewidth]{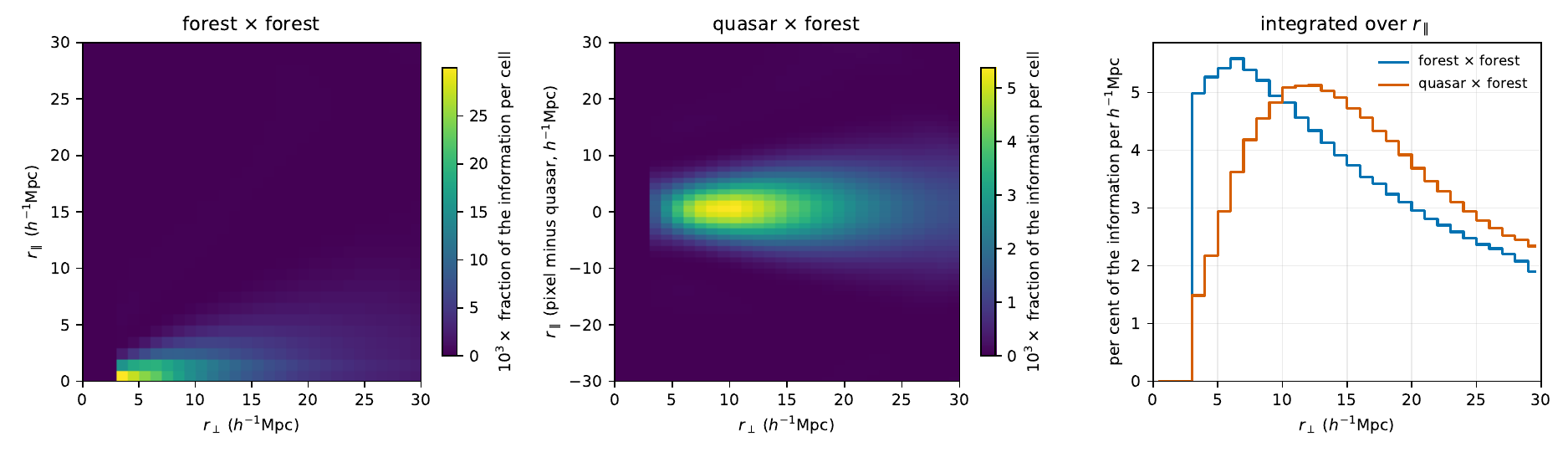}
\caption{Where the lensing information comes from: the fraction of the unmarginalised science response in each $1\Mpch$ cell of $(\rperp,\rpar)$, for the forest auto-correlation (left) and the quasar--forest cross-correlation (middle), accumulated over the DR1 pair catalogs with the fiducial combined template, and (right) the same densities integrated over $\rpar$, i.e.\ the information per unit $\rperp$. Pairs with $\rperp<3\Mpch$ are excluded. }
\label{fig:scales}
\end{figure*}

Which pair separations carry the lensing information? We use the contribution of each pair to the unmarginalised science response as a diagnostic of scale sensitivity. In the independent-pair Gaussian approximation, and before fitting nuisance amplitudes, this corresponds to the Fisher-information density. Figure~\ref{fig:scales} shows this density accumulated in $1\Mpch$ cells of $(\rperp,\rpar)$ over the production pair catalogues of both statistics, with the production weights, cuts and kernels and the fiducial combined template summed over its five science bands; for the quasar--forest pairs the weight is $w_p$ and $\rpar$ is signed.

We see that the majority of the signal comes from small $\rpar$ separations and relatively small $\rperp$ separations. Integrated over $\rpar$, the information per unit $\rperp$ peaks at $\rperp\sim7\Mpch$ for the forest--forest and at $\rperp\sim13\Mpch$ for the quasar--forest correlation, with half of it inside $13$ and $16\Mpch$ respectively.

\subsection{Error derivation}

The measurement errors are obtained by jackknife: the sums of Equation~\ref{eq:qe} are accumulated per HEALPix region of the pair midpoints ($\Nside=8$, $\sim 300$ regions on the DR1 footprint), the estimator is re-run leaving one region out at a time, and the covariance of the amplitudes, including that between the two statistics and between the sub-fits below, is the scaled scatter of the leave-one-out estimates.

To check the error scale, we replace the combined template by 100 independent Gaussian random realisations of its spectrum through the tracer mask, holding the forest data fixed. For simplicity, this diagnostic fits eleven combined-template components, not the full 55-component slice model. The mean amplitudes are $-0.022\pm0.045$ (auto) and $-0.051\pm0.061$ (cross), where the errors here are standard errors of the Monte Carlo means. The per-realisation scatters are $0.447$ and $0.612$, compared with root-mean-square jackknife errors of $0.445$ and $0.587$, close to the jack-knife error estimates (from a single realization).

\subsection{The Response Matrix}

\begin{figure}
\centering
%% To reproduce: python scripts/joint_response_fit.py --auto $LYALENSER_DATA/stageb/dr1_lowz_v8 --cross $LYALENSER_DATA/stageb/dr1_qso_v2 --lowz $LYALENSER_DATA/lowz_v5 --bands 40 200 400 600 800 1000 --tag v5; then python scripts/joint_response_fit.py --plot-only v5 auto (LyaLenser repository) -> report/figures/response_matrix_auto_v5.pdf, copied here
\includegraphics[width=\linewidth]{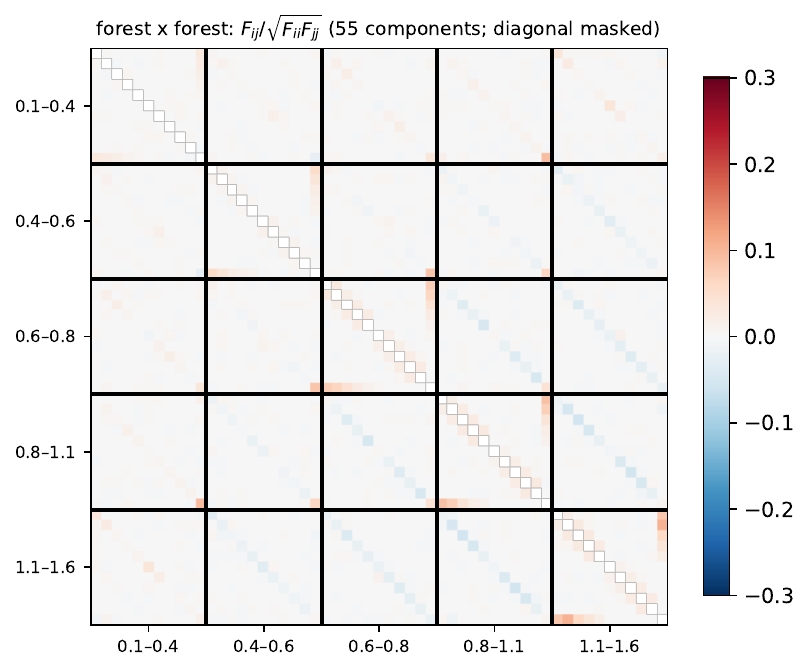}
\caption{The joint response matrix $F_{ij}/\sqrt{F_{ii}F_{jj}}$ of the forest auto-correlation: 55 components each: for every tracer slice (thick lines) the five science bands, their five curl partners and the junk band.  The quasar--forest cross-correlation looks visually identical. }
\label{fig:response}
\end{figure}

We start with the response matrix. All slice templates enter one fit: with five slices, five bands, their curl partners and the junk band, $F$ has $55\times55$ elements for each statistic, accumulated per jackknife region. Figure~\ref{fig:response} shows it normalised to unit diagonal. It is close to diagonal: in addition to expected small neighboring $\ell$ region correlations, 
the largest off-diagonal elements, about $0.11$, connect the junk component to the $200$--$400$ band in the highest-redshift slice; between redshift slices their magnitudes remain below $0.10$.

This means that the results are largely independent. We could fit any one slice and get a largely independent answer.

\subsection{The lensing amplitude}
\label{sec:amplitude}
Finally, we can look at the results of our fiducial analysis. Putting everything together we find
\begin{align}
A_L&=0.42\pm0.43\quad\text{(forest$\times$forest)},\nonumber\\
A_L&=0.87\pm0.48\quad\text{(quasar$\times$forest)},\nonumber\\
A_L&=0.62\pm0.33\quad\text{(combined)}.\nonumber
\end{align}
The two statistics are essentially independent due to noise domination , so the combination is close to inverse-variance, with the cross-correlation carrying 44 per cent of the weight. The joint curl amplitudes are $-0.47\pm0.59$ and $-0.62\pm0.79$. The result is consistent with the $\Lambda$CDM expectation of one at $1.2\sigma$ and with no lensing at $1.9\sigma$.

\begin{figure}
\centering
%% To reproduce: python scripts/paper_figures.py --split (LyaLenser repository; reads results/auto_cross_combination_v5.json, built by combine_auto_cross.py from dr1_lowz_v8 and dr1_qso_v2 with the sub-slab fits)
\includegraphics[width=\linewidth]{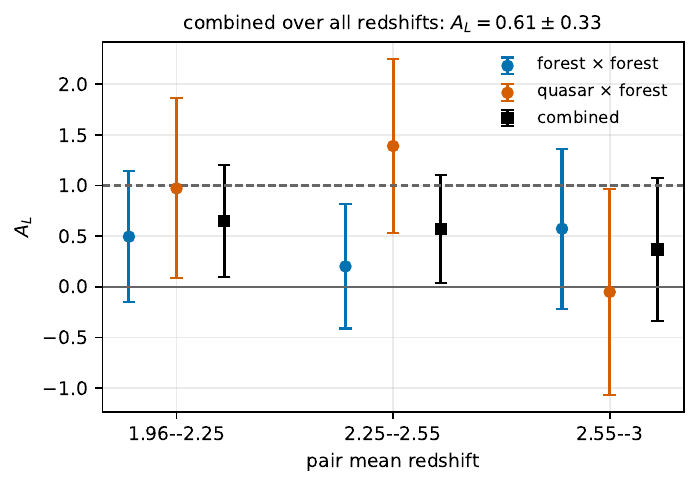}
\caption{The lensing amplitude in three ranges of the pair mean redshift, for the forest auto-correlation, the quasar--forest cross-correlation and their optimal combination (jackknife errors); the dashed line is the $\Lambda$CDM value.}
\label{fig:zsplit}
\end{figure}

Figure~\ref{fig:zsplit} shows the amplitudes split by the mean redshift of the pairs (three ranges of the forest, each fitted with the combined template), for the two statistics and their combination: the split is flat within its errors.

\begin{figure}
\centering
%% To reproduce: python scripts/paper_figures.py --slices (LyaLenser repository; reads results/joint_fit_v5.json)
\includegraphics[width=\linewidth]{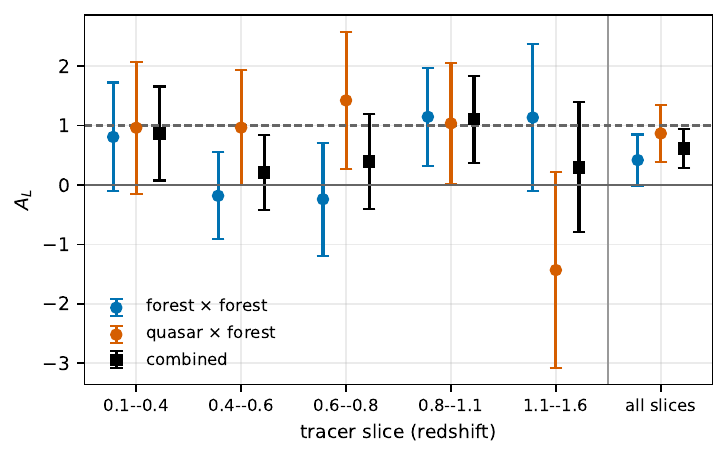}
\caption{The lensing amplitude fitted to each tracer slice on its own within the joint fit, for the forest auto-correlation, the quasar--forest cross-correlation and their combination (jackknife errors), and the fit to all slices at the right; the dashed line is the $\Lambda$CDM value.}
\label{fig:ssplit}
\end{figure}

Figure~\ref{fig:ssplit} shows the amplitudes fitted to the individual tracer slices, for the two statistics and their combination, next to the fit to all slices. Every slice is consistent with the $\Lambda$CDM value and with the others. No slice dominates: the first three, which carry two thirds of the convergence power of Figure~\ref{fig:kernel}, also carry two thirds of the weight.

\begin{figure}
\centering
%% To reproduce: python scripts/paper_figures.py --bands (LyaLenser repository; reads results/joint_fit_v5.json)
\includegraphics[width=\linewidth]{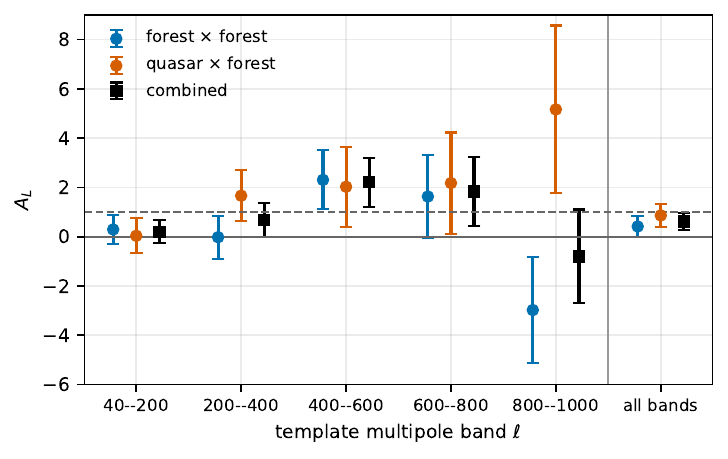}
\caption{The lensing amplitude fitted to each science band of the templates within the joint fit, for the forest auto-correlation, the quasar--forest cross-correlation and their combination (jackknife errors), and the fit to all bands at the right; the dashed line is the $\Lambda$CDM value.}
\label{fig:bsplit}
\end{figure}

Figure~\ref{fig:bsplit} completes the splits with the amplitudes per template multipole band. Again we find that each individual band is consistent with $\Lambda$CDM and that no band dominates, although the greatest weight is given to the largest scales. 

\begin{figure}
\centering
%% To reproduce: python scripts/paper_figures.py --robustness (LyaLenser repository). Rows read results/joint_fit_v5.json (fiducial, the two statistics, block-diagonal R), auto_cross_combination_{l1300,l500,rp40,rp20,kmed,kfine}_v5.json, each built by combine_auto_cross.py from the dr1_lowz_<tag>_v5 and dr1_qso_<tag>_v5 products of slurm/refit.sbatch (the variant catalogues of slurm/robustness.sbatch refitted with the derivative-map templates), and joint_fit_v5_noderiv.json (joint_response_fit.py --no-derivative).
\includegraphics[width=\linewidth]{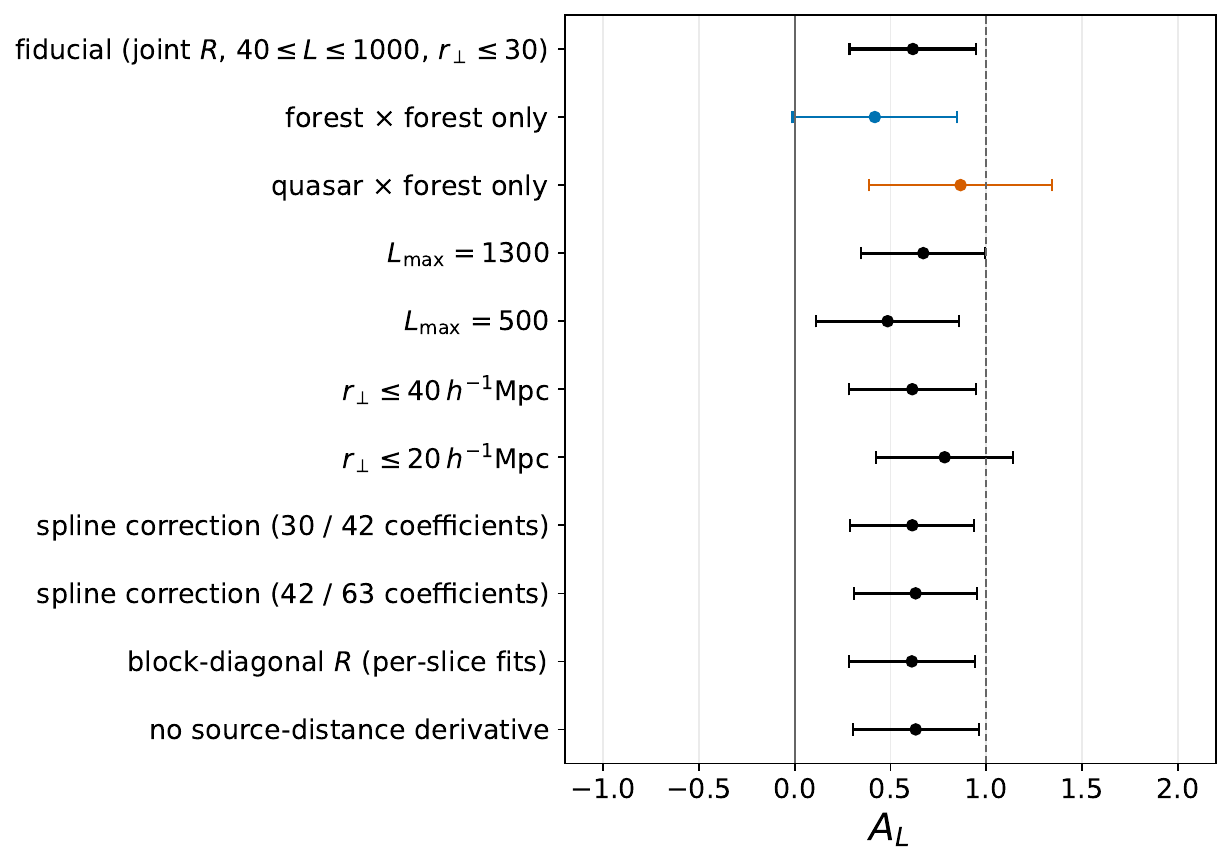}
\caption{The lensing amplitude under the analysis variants described in the text, one-sigma jackknife errors; the solid and dashed lines mark $A_L=0$ and the $\Lambda$CDM value. The first row is the fiducial combination and the next two its two statistics; the remaining rows are combinations with the science window widened to $L=1300$ or narrowed to $L=500$, the transverse pair range widened to $40\Mpch$ or narrowed to $20\Mpch$, the spline correction of both correlation functions refitted with more knots (medium: 30 coefficients for the forest and 42 for the quasar--forest correction; fine: 42 and 63, against 16 and 20 in the fiducial), and the response matrix restricted to its per-slice blocks (the variant rows use per-slice response matrices, so this row is their reference); the last row is the joint fit without the source-distance derivative of the templates.}
\label{fig:robust}
\end{figure}

Figure~\ref{fig:robust} shows how the result responds to the main analysis choices. Each variant re-runs both statistics from the pair catalogue with one choice changed and combines them as in the fiducial.  Spline correction entries change the number of points used to fit the correlation function (Section \ref{sec:kernel}). The last row  switches off the source-distance derivative of the templates (Section~\ref{sec:sourcedist}).  We see that results are remarkably stable with respect to the choice of analysis parameters.

%Widening the science window to $L_{\max}=1300$ (a sixth band $1000$--$1300$, templates rebuilt to that band limit) gives $0.67\pm0.32$, no gain in precision because the added band is noise-dominated and its tracer bias an extrapolation; narrowing it to $L_{\max}=500$ (five bands of width 60--100) gives $0.49\pm0.38$, i.e.\ the bands above 500 contribute a fifth of the statistical weight and pull the amplitude up by $0.13$. Changing the transverse range of the pairs, which changes which part of the correlation function sets the response, moves the amplitude within its error: $0.61\pm0.33$ for $r_\perp\le40\Mpch$ and $0.78\pm0.36$ for $r_\perp\le20\Mpch$. Restricting the response matrix to its per-slice blocks gives $0.61\pm0.33$. All variants are within $0.5\sigma$ of the fiducial, and the shifts are of the size expected from the noise they do not share.

\section{Conclusions}
\label{sec:conclusions}
We have made a first attempt to measure the weak lensing of the \lya forest on the DESI DR1 forest sample, by projecting a quadratic estimator of the forest pair correlations onto deflection templates built from the low-redshift large-scale structure. The estimator is the pair-based analogue of the CMB lensing quadratic estimator: the local departure of the forest correlation from its sky mean, weighted by the transverse derivative of the fitted correlation function, is summed against the deflection expected from the tracers. We applied it to two statistics, the forest auto-correlation and the quasar--forest cross-correlation.
The templates combine eleven tracer samples from DESI DR1 and BOSS in five redshift slices, with weights measured from their full set of auto- and cross-spectra. The templates are validated against CMB lensing, the ACT DR6 and Planck PR4 convergence maps recovering $0.99\pm0.04$ and $0.93\pm0.03$ of the predicted cross-correlation, and the estimator itself by injecting the template deflection into the data.
 
The result is $A_L=0.42\pm0.43$ from the forest auto-correlation, $0.87\pm0.48$ from the quasar--forest cross-correlation and $A_L=0.62\pm0.33$ combined, with all errors from a jackknife over about 300 sky regions.  

At this level of precision, the noise realization plays an important role in detection significance. We could have measured $A_L = 1.4\pm0.33$ and claimed a four sigma detection since 1.4 is a normal up-scatter from one but would be a very unlikely as an up-scatter from zero. We could have measured $A_L = 1.02\pm0.33$ and claim a 3 sigma detection. Perhaps we could measure  $0.71\pm0.33$ and desperately claim a two sigma preference for lensing over null. Instead we measured $A_L = 0.62\pm0.33$. It is consistent with $A_L=1$. Bastards.

 The curl amplitudes and the random-template nulls are consistent with zero, and the result is stable against the width of the science window, the transverse range of the pairs and the treatment of the response matrix (Figure~\ref{fig:robust}). The precision is close to what the forecasts for a DR1-sized sample anticipated \citep{1706.07870,1706.08939} for the forest-forest correlations.

The cross-correlation, which we introduced here, carries almost half of the statistical weight although it uses the same forest pixels.  This makes the prospects for the complete DESI sample favourable. The five-year survey will roughly double the sightline and quasar densities over the same footprint and the tracer catalogues used for the templates will be complete over it. At fixed area the auto-correlation error scales inversely with the sightline density and the cross-correlation error with the inverse square root of the number of quasar--sightline pairs, so we expect the combined error to shrink by a factor of two to three, to a 10--20 per cent measurement of the lensing amplitude. This would then provide a new and systematically independent data-point in the growth history of the Universe.

\section*{Acknowledgements}

AS acknowledges useful discussions with Prakruth Adari who spear-headed an earlier iteration of this project in the bronze age, Andreu Font, Andrei Cucei and Rupert Croft.

\section*{Statement on AI use}
AI has been heavily used in the development of this work. All the code has been written and executed by agents, including analysis and plotting; mostly by Claude Fable with adversarial review by Codex Astra under close guidance by the author.

The paper narrative has been largely human-written. The main analysis code was completed in a week, the remaining several weeks were spent on human understanding and tweaking of the code, analysis parameters and tests and rewriting the AI generated report into an article that is human cosmologist readable. Large parts of analysis, including additional mock tests were omitted from the write-up to create a self-contained structured piece of work. The full repository is publicly available at \url{https://github.com/slosar/LyaLenser}.

\appendix

\section{Wiener filtering cross-correlation with the truth}
\label{app:wiener}
This is a standard piece of liner algebra lore that makes matched filter that use Wiener filtered templates automatically unbiased in amplitude.
\newcommand{\vs}{\mathbf{s}}
\newcommand{\vn}{\mathbf{n}}
\newcommand{\vd}{\mathbf{d}}
\newcommand{\vw}{\mathbf{w}}

Consider a general case with  signal $\vd = \vs + \vn$ with signal covariance $\left<\vs \vs\right>^T=S$ and noise covariance $\left<\vn \vn^T\right>=N$ and we also make the standard assumption that signal and noise are uncorrelated so $\left<\vd \vd^T\right>=S+N$.
The Wiener filter is defined as signal to total power weighting so $\vw = S (S+N)^{-1} \vd$. We have
\begin{eqnarray}
\left<\vw\vs^T\right> &=& S (S+N)^{-1} S \\
\left<\vw\vw^T\right> &=& S (S+N)^{-1} (S+N) (S+N)^{-1}S \nonumber \\&=& S (S+N)^{-1} S  
\end{eqnarray}
so $\left<\vw\vs^T\right> = \left<\vw\vw^T\right>$. The covariance of Wiener filter with itself is the same as covariance of wiener filter with the truth. Under these ideal Wiener assumptions, the ensemble response ratio is $\left<\alpha^T\alpha^{\rm truth}\right>/\left<\alpha^T\alpha^T\right>=1$.

\section{Effects of kernel error}
\label{app:kernel_error}
The response kernel $g=\xi_F'$ is the derivative of a correlation function fitted to noisy measured cells, so the kernel we use is $\hat g=g+\delta g$, with $g$ the true derivative and $\delta g$ the error of the fit. With the pair weights $W$ written explicitly, the amplitude estimator of Equation~\ref{eq:qe} for a single template is
\begin{equation}
\hat A=\frac{\sum W\,G\,(\dF\dF-\hat\xi_F)\,\hat g}{\sum W\,G^2\,\hat g^2}.
\end{equation}
The lensing signal in a pair is $\langle\dF\dF-\xi_F\rangle=A\,G\,g$, so the numerator averages to $A\sum WG^2(g^2+g\,\delta g)$, while the denominator, the response, is $\sum WG^2(g^2+2g\,\delta g+\delta g^2)$. Approximating the expectation of the ratio by the ratio of expectations (good when $\delta g/g \ll 1$), the quadratic term gives an attenuation estimate:
\begin{equation}
\langle\hat A\rangle\simeq A\left[1-\frac{\langle WG^2\,{\rm Var}(\delta g)\rangle}{\langle WG^2 g^2\rangle}\right].
\label{eq:attenuation}
\end{equation}

This illustrates how kernel uncertainty can bias the normalization as well as add noise. At fixed correlation measurement, increasing the estimator's pair count alone does not remove it; improving the correlation measurement can reduce ${\rm Var}(\delta g)$. Since the kernel is fitted from the same data, neglected covariance terms can also matter.

%Both sides of the trade-off can be measured. Adding free parameters to the correction $S$ removes misfit, which is a systematic error in $g$ of the other kind, and adds fitted noise, which the derivative amplifies since neighbouring cells are only $1\Mpch$ apart. The gain is the change of the amplitude on the data between kernels, which tracks the misfit and saturates almost at once: relative to the two-parameter physical kernel the amplitude is $1.077$ with one cubic segment in each direction (16 coefficients for $S$ plus 6 for $N$), $1.099$ with 30 and $1.100$ with 42. The cost is the paired response on simulated forest realisations with a known injected deflection, one minus the attenuation of Equation~\ref{eq:attenuation}, which does not saturate: it falls by $1.4\pm1.2$, $4.0\pm1.8$ and $10.4\pm2.3$ per cent for the same three kernels. We therefore adopt the smallest surface that removes the misfit, the bicubic in $(\rperp,\rpar^2)$ with 22 parameters, whose cost is not measurable at our precision. With a larger sample the term $\langle WG^2\,{\rm Var}(\delta g)\rangle$ could be estimated from the covariance of the measured cells and subtracted from the response matrix, which would remove the bias for any number of parameters.

\bibliography{references.bib}% Produces the bibliography via BibTeX.

\end{document}